\documentclass[12pt,preprint]{aastex}
\shorttitle{{\it Spitzer} Observations of High Redshift QSOs}
\shortauthors{Hines et al.}

\begin{document}

%% LaTeX will automatically break titles if they run longer than
%% one line. However, you may use \\ to force a line break if
%% you desire.

\title{{\it Spitzer} Observations of High Redshift QSOs}

%% Use \author, \affil, and the \and command to format
%% author and affiliation information.
%% Note that \email has replaced the old \authoremail command
%% from AASTeX v4.0. You can use \email to mark an email address
%% anywhere in the paper, not just in the front matter.
%% As in the title, use \\ to force line breaks.

\author{Dean C. Hines\altaffilmark{1}, O. Krause\altaffilmark{2}, G.
H. Rieke\altaffilmark{2}, X. Fan\altaffilmark{2}, M.
Blaylock\altaffilmark{2}, G. Neugebauer\altaffilmark{2}}

%% Notice that each of these authors has alternate affiliations, which
%% are identified by the \altaffilmark after each name.  Specify alternate
%% affiliation information with \altaffiltext, with one command per each
%% affiliation.

\altaffiltext{1}{Space Science Institute, 4750 Walnut Street,
Suite 205 Boulder, CO 80301}
\altaffiltext{2}{Steward Observatory, The University of Arizona, 933 
N. Cherry Ave., Tucson, AZ 85721}

%\altaffiltext{3}{Spitzer Science Center, California
%Institute of Technology, MS 220-6, Pasadena, CA 91125;}

%% Mark off your abstract in the ``abstract'' environment. In the manuscript
%% style, abstract will output a Received/Accepted line after the
%% title and affiliation information. No date will appear since the author
%% does not have this information. The dates will be filled in by the
%% editorial office after submission.

\begin{abstract}

We have observed 13 z $\ge$ 4.5 QSOs using the Multiband Imaging
Photometer for {\it Spitzer}, nine of which were also observed with
the Infrared Array Camera.  The observations probe rest wavelengths
$\sim 0.6-4.3\mu$m, bracketing the local minimum in QSO spectral
energy distributions (SEDs) between strong optical emission associated
directly with accretion processes and thermal emission from hot dust
heated by the central engine.  The new {\it Spitzer} photometry
combined with existing measurements at other wavelengths shows that
the SEDs of high redshift QSOs (z $\ge$ 4.5) do not differ
significantly from typical QSOs of similar luminosity at lower
redshifts ($z \lesssim 2$).  This behavior supports other indications
that all the emission components and physical structures that
characterize QSO activity can be established by $z = 6.4$.  The
similarity also suggests that some QSOs at high redshift will be very
difficult to identify because they are viewed along dust-obscured
sight lines.

\end{abstract}

\keywords{ infrared: galaxies --- quasars: general --- galaxies:
high-redshift}

\section{Introduction}

Recent optical surveys have revealed many QSOs with redshifts z $>$
4.5, including several with z $>$ 6.  High resolution imaging
indicates that their luminosities are intrinsic and not caused by
lensing (e.g., Fan et al.  2001, 2003; Richards et al.  2004).  X-ray
(Brandt et al.  2002; Mathur, Wilkes \& Ghosh 2002; Shemmer et al.
2005), UV/optical and near-IR spectroscopic observations (Fan et al.
2003; Barth et al.  2003; Maiolino et al.  2003) show that these
sources share many properties with luminous QSOs at lower redshift ($z
\lesssim 2$).  In particular, ultraviolet continua, high ionization
emission lines, and strong low ionization MgII and FeII emission are
all present in proportions consistent with the low redshift
counterparts.

Submillimeter and centimeter radio observations indicate that these
QSOs have copious dust (e.g., Omont et al.  1996; Bertoldi et al.
2003a, 2003b; Robson et al.  2004; Carilli et al.  2004), but this
cool material may simply be heated by massive star formation in highly
luminous host galaxies, and thus be unrelated to the QSO activity.
Active nuclei can heat dust directly to produce stronger mid-infrared
emission than is characteristic of star formation.  This behavior is
common in radio galaxies and QSOs at low redshift (Neugebauer et al.
1986; Low et al.  1988, 1989; Polletta et al.  2000; Wilkes et al.
2000; Kuraszkiewicz et al.  2003; Haas et al.  1998, 2000, 2003; Shi
et al.  2005).  However, the {\it IRAS} and {\it ISO} observatories
were not sensitive enough to probe this type of emission in
high-redshift QSOs.

The {\it Spitzer Space Telescope (Spitzer}: Werner et al.  2004)
provides sufficient sensitivity to search for this mid-IR emission
from even the most highly redshifted QSOs currently known.  As part of
a large survey of luminous AGN, we have observed 13 high redshift (z
$>$ 4.5), UV/optically-selected QSOs with the goal of comparing their
spectral energy distributions (SEDs) with lower redshift objects.  In
this Letter, we use the new observations to study the rest wavelengths
between 0.6 and 4.3$\mu$m, a region containing the characteristic
local minimum ($\lambda \sim 1.5\mu$m) in QSO SEDs as well as the
thermal emission from the hot (T$_{\rm dust} \gtrsim 300$K),
AGN-heated dust.  A complementary General Observer program will report
on a larger sample of high redshift QSOs and will investigate their
SEDs from X-ray through the {\it Spitzer} and radio wavelengths (Fan
et al., in prep).

\section{Observations}

Images at 24$\mu$m ($5\farcs25 \times 5\farcs25$/pixel) were obtained
for each object using the photometry mode of MIPS (Rieke et al.  2004)
as part of the MIPS GTO program titled ``The Far-IR Spectral Energy
Distributions of Luminous Active Galactic Nuclei'' (PID 82).  After
initial processing by the {\it Spitzer} Science Center (SSC) pipeline
(S10.5.0) to provide reconstructed pointing information, the MIPS data
were further processed into a final image using the MIPS Data Analysis
Tool (DAT, ver.  2.71) developed by the MIPS Instrument Team (Gordon
et al.  2004, 2005).  Aperture photometry was performed on the images
with IDP3 (version 2.9: Schneider \& Stobie 2002), using a 14\farcs7
target aperture and a background annulus from 29\farcs4 -- 41\farcs7.
The median background per pixel was scaled to the appropriate target
aperture size and subtracted from the summed target aperture flux for
each object.  The final flux in instrumental units was then corrected
to an infinite aperture, and conversion to physical flux density units
was performed by multiplying the instrumental total flux by the MIPS
calibration factor (MIPS Data Handbook, V3.1, hereafter MDH3.1).
Random uncertainties in the background--subtracted estimates of
on--source flux were estimated from the {\it rms} pixel-to-pixel
dispersion inside the background annulus measured on the image and
scaled to the area of the target aperture.  The final uncertainties
are dominated by the absolute calibration uncertainty of 10\%.

Infrared Array Camera (IRAC: Fazio et al.  2004) observations of nine
QSOs were obtained in channels 1, 2, 3 and 4 (3.6, 4.5, 5.8 and
8.0$\mu$m).  Images were obtained as Basic Calibrated Data products
from the SSC data pipeline (S11.5.0) as described in the SOM4.6 and
the Pipeline Description Document available through the SSC. Aperture
photometry was performed using a 3 pixel (3\farcs66) target aperture
and background annuli from 3-7 pixels.  Other aspects of the
photometry were similar to the procedures used at 24$\mu$m.  The
measurement uncertainties were estimated from the {\it rms} noise in
the background annulus.  The uncertainty in absolute calibration is $3
- 5\%$ (Reach et al.  2005).

The measured flux densities for each object are presented in Table 1
along with their measurement uncertainties.  The QSOs were detected
with high signal-to-noise in the MIPS 24$\mu$m band, and in all four
bands for the nine objects that were observed with IRAC. The objects
were unresolved, except for BR 1202-0725, which is known to have a
nearby, IR-bright companion galaxy 2\farcs6\ to the NW (e.g., Hu et
al.  1996; Ohyama et al.  2004).  For this object, the photometry in
the IRAC and MIPS bands was re-derived using a two component PSF-fit.
Only the flux densities attributable to the QSO are presented in Table 1.

\section{Spectral Energy Distributions}

Figure 1 shows the rest-frame UV through mid-IR SEDs of the high
redshift sample.  A composite SED for radio-quiet QSOs is also shown
in each panel for comparison (Elvis et al.  1994).  The composite
represents the average behavior of low redshift QSOs, but individual
objects can show significant deviations.  In Figure 1c, we also
present the radio loud composite SED from Elvis et al.  (1994), and a
composite constructed only from the PG QSOs (Sanders et al.  1989).
The high redshift QSOs generally match the composite SEDs and would
not be obviously distinguishable from typical, optically selected
luminous QSOs at low redshift.  In particular, the rise in the
high-redshift SEDs for $\lambda \gtrsim 1.5\mu$m is likely to be the
signature of thermal emission from hot dust (e.g., Rieke \& Lebofsky
1981; Cutri et al.  1985; Barvanis 1990; Elvis et al.  1994; Haas et
al.  1998, 2000, 2003).

The deviations from the composite SEDs are not surprising.  A few
objects (e.g., SDSS J1148+5251, BR0004-6224) show a more rapid rise
into the infrared, and SDSS J0836+0054 is radio loud (Petric et al.
2003), but their behavior still lies within the scatter seen for lower
redshift objects (Elvis et al.  1994).  The rapid decrease for
$\lambda \lesssim 0.2\mu$m in the SEDs of the three QSOs with z$\ge
5.8$ is caused by absorption from the Ly$\alpha$ forest in the
observed frame.  In addition, the photometric bands that contain the
H$\alpha$ emission line in each object are elevated above the 
extrapolated continuum.

At low redshift, the strength of the $\lambda \approx 3.7\mu$m
infrared band is correlated with the $B$-band flux density implying
that IR emission arises from hot dust heated directly by the
UV/optical continuum from the central active nucleus (Rieke \&
Lebofsky 1981; Cutri et al.  1985; Barvanis 1990; Haas et al.  2003).
The rest frame flux density ratios F$_{\nu}(3.7\mu$m)/F$_{\nu}$(B)
derived from the three composite SEDs shown in Figure 1c are 7.8, 6.0
and 5.6 for the radio-quiet, radio-loud QSOs and the PG QSOs,
respectively.  A recent {\it Spitzer} investigation of the SEDs of
unreddened, low-redshift QSOs identified in the Sloan Digital Sky
Survey (SDSS) suggests that this ratio can be somewhat larger ($\sim
8-9$: Richards et al.  2006).  The ratio for our sample of high
redshift QSOs is $8.1\pm1$.  X-ray to UV/optical emission is also
correlated, with the X-ray to UV/optical spectral index depending on
the luminosity of the AGN, but there is negligible dependence of the
spectral index on redshift (e.g., Strateva et al.  2005).  These
correlations together suggest that the source of the 3.7$\mu$m
emission in QSOs is the same over a range of luminosities and for
redshifts from z$\sim$ 0.1 to 6.4, and that the mid-infrared feature
is emitted from dust heated directly by the central active nucleus.

\section{Discussion}

Many lines of evidence suggest that the high-redshift QSOs are
essentially indistinguishable from their lower redshift counterparts.
The rest frame UV/optical continua and emission lines (e.g., Barth
et al.  2003; Fan et al.  2004), and the derived metallicities, are
all typical of low redshift QSOs (e.g., Pentericci et al.  2002;
Maiolino et al.  2003).  The objects also have similar X-ray (Brandt
et al.  2002; Mathur, Wilkes \& Ghosh 2002), sub-millimeter (Bertoldi
et al.  2003a; Robson et al.  2004), and radio properties (Bertoldi et
al.  2003b; Carilli et al.  2004; Frey et al.  2005).  Estimates of
the masses of the super-massive black holes in some of these
high-redshift systems imply M$_{\rm SMBH} \sim 10^{9}$M$_{\odot}$
(e.g., Willott, McLure \& Jarvis 2003), comparable to low redshift
QSOs.  Our {\it Spitzer} measurements strengthen the evidence that
QSOs have similar properties over the entire range of observed
redshifts.  Thus, the fundamental structures that characterize
luminous QSOs in the local universe, including hot dust heated
directly by the active nucleus, were already in place in these
objects $\sim 900$ Myrs after the big bang\footnote{H$_{\rm 0} = 75$ 
km s$^{-1}$ Mpc$^{-1}$, $\Omega_{M} = 0.3$, and $\Omega_{\Lambda} = 
0.7$.}.

The presence of substantial amounts of both cold and hot dust around
the high-redshift QSOs (e.g., Bertoldi et al.  2003a; Robson et al.
2004; this work) is consistent with their apparently unobscured
ultraviolet continua only if, as with lower-redshift QSOs (e.g.,
Neugebauer et al.  1986; Low et al.  1988, 1989), the dust is in
structures that leave some lines of sight unobscured.  Many of the
ultra- and hyperluminous infrared galaxies have been shown to harbor
QSOs that are typical of UV/optically-selected objects, but are
obscured from direct view by dust (e.g., Hines et al.  1995; 1999;
Young et al.  1996; Goodrich et al.  1996; Tran et al.  2000).  The
many so-called Type-2 QSOs that have been identified (at $z \le 1$) in
the SDSS (Zakamska et al.  2003, 2004, 2005) and with {\it Spitzer}
(e.g., Mart{\'{\i}}nez-Sansigre et al.  2005) also must have obscuring
material.  Possibilities for these structures include the
circumnuclear torus envisioned in simple unified schemes (e.g.,
Antonucci 1993; Rowan-Robinson 1995; Urry \& Padovani 2000) or more
complex structures in disk winds (e.g., Konigl \& Kartje 1994; Kartje
\& Konigl 1996; Elvis 2000).

In this context, the currently identified high redshift QSOs would
then represent only a fraction of the true space density of such
objects, with the remainder obscured in our direction by their
circumnuclear material.  Furthermore, reionization of the
intergalactic medium by these objects would not be distributed
spherically around each QSO, but would instead extend in reionization
(bi) cones analogous to the ionization bicones seen in the
interstellar media of nearby QSOs.

It appears that the solid state emission features associated with
aromatic molecules play an important role in the {\it Spitzer}
24$\mu$m detections of objects out to $z \sim 2.2$ (e.g., Caputi et
al.  2005).  At larger redshifts, the effects of these features on the
detection rates will decrease, because: 1.)  they begin to move out of
the MIPS band; 2.)  at the extremely high galaxy luminosities required
for detections, the equivalent widths of the features are reduced; and
3.)  the effects of low metallicity and hard UV continua will tend to
destroy the carriers (e.g., Englebracht et al.  2005 and references
therein).  Our results show that the rest-frame near- and mid-infrared
continua of luminous AGNs remain strong to at least $z \sim 6$.  Thus,
we expect that such AGNs will play an increasing role in the 24$\mu$m
detections above $z = 2.2$, and will probably dominate the detections
for $z > 3$.

\section{Conclusion}

    From new 3.5, 4.5, 5.8, 8.0 and 24$\mu$m {\it Spitzer} measurements
    of high-z QSOs ($z \ge 4.5$) we find:

\begin{itemize}

\item The QSOs are easily detected with IRAC and MIPS at 0.6-4.3$\mu$m
in the rest frame.

\item There is no obvious evolutionary trend in the UV through mid-IR
SEDs compared with low redshift QSOs ($z \lesssim 2$).

\item The high redshift QSOs conform to the relationship between
L$_{B}$ and L$_{3.7\mu\rm m}$ as established for low redshift AGN,
supporting the conclusion that the rest-frame emission near 4$\mu$m in
the high redshift objects originates from hot dust heated directly by
the central active nucleus.

\item AGNs will play an increasing role in the {\it Spitzer} 24$\mu$m
detections above $z = 2.2$, and will probably dominate the detections
for $z > 3$.

\end{itemize}

Our results add further evidence that the fundamental structures of
QSOs can be established by $z = 6.4$, corresponding to $\sim 900$ Myr
after the big bang.  They also suggest that there is a population of
QSOs at these redshifts that will be extremely difficult to identify
due to dust obscuration along the line of sight.

\acknowledgments

We thank D. Sanders and B. Wilkes for supplying electronic versions of
the composite QSO SEDs from Sanders et al.  (1989) and Elvis et al.
(1994).  We also thank M. Brotherton and B. Wills for comments that
contributed significantly to the clarity of the manuscript.  This
publication makes use of data products from the Two Micron All Sky
Survey, which is a joint project of the University of Massachusetts
and the Infrared Processing and Analysis Center/California Institute
of Technology, funded by the National Aeronautics and Space
Administration and the National Science Foundation.  We have used the
NED and SIMBAD databases.  This work is based [in part] on
observations made with the Spitzer Space Telescope, which is operated
by the Jet Propulsion Laboratory, California Institute of Technology
under NASA contract 1407.  Support was provided by NASA through
contract 1255094 issued by JPL/Caltech.

\clearpage

\begin{figure}
\epsscale{.80}
%\plotone{f1.pdf}
\plotone{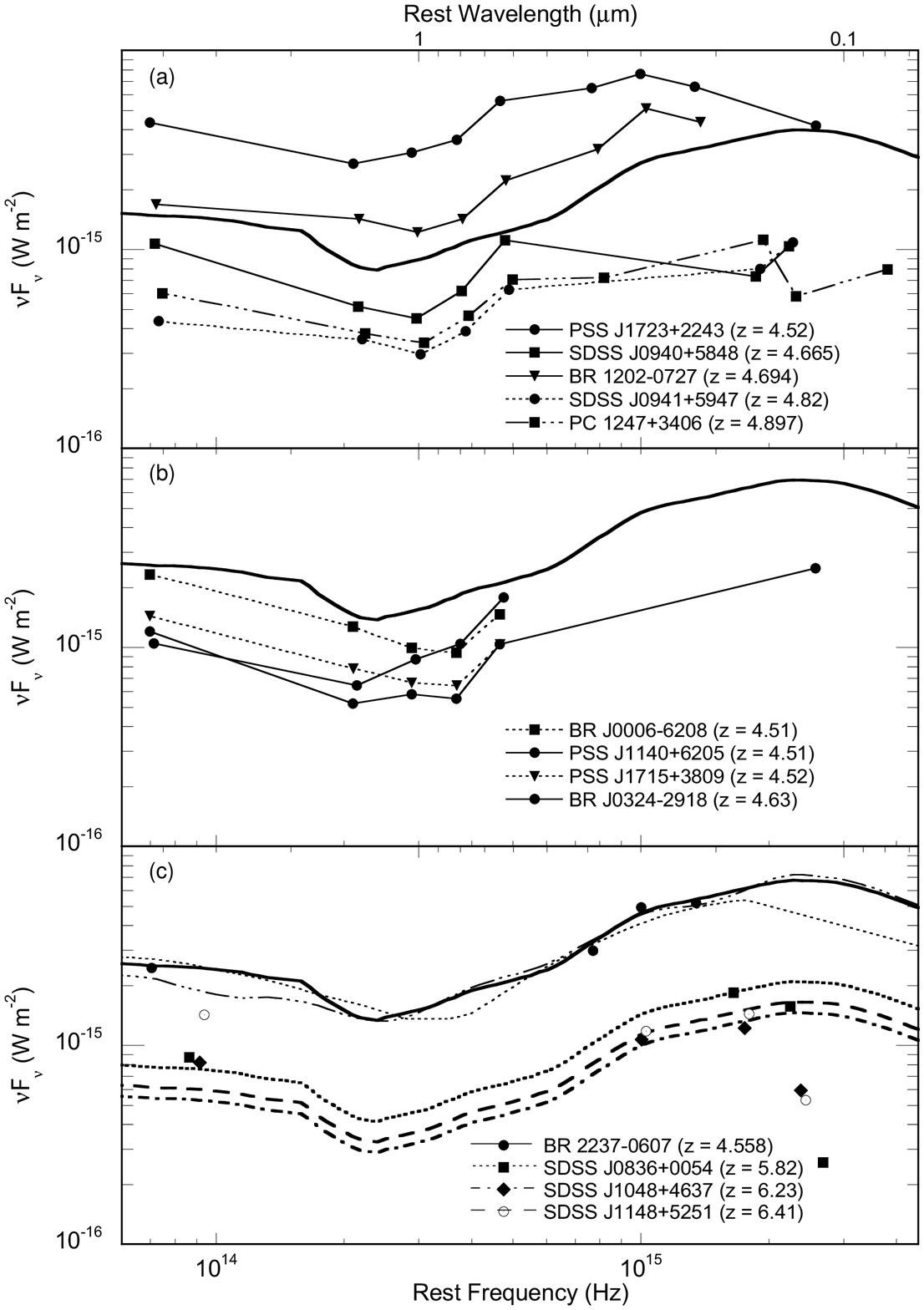}
  \caption{\footnotesize Rest-Frame Spectral Energy Distributions
  (SEDs) of high-redshift QSOs observed by {\it Spitzer}.
  Ground-based optical and near-IR data are also included for most
  objects.  The objects and redshifts are indicated in the legends.
  Panels (a) \& (b) show objects for which both MIPS and IRAC
  observations were obtained.  Heavy solid curves (without individual
  data points) in panels (a) \& (b) show the Elvis et al.  (1994)
  radio-quiet QSO composite SED for comparison.  Panel (c) shows
  objects for which only MIPS 24$\mu$m observations were obtained.
  Panel (c) also presents the Elvis et al.  (1994) radio-quiet
  composite SED, but in this panel the SED is fitted to the rest frame
  UV/optical photometry for each object.  For comparison, BR 2237-0607
  is also fit with: (dotted line) the radio-loud composite SED of
  Elvis et al.  (1994); and (triple-dot, long-dash line) a composite
  SED constructed from PG QSOs (Sanders et al.  1989).}
\end{figure}

\clearpage

\begin{deluxetable}{llccccccccccccccc}
\tabletypesize{\scriptsize}
\rotate
\tablecaption{{\it Spitzer} Observed Photometry of High Redshift QSOs}
\tablewidth{0pt}
\tablehead{
\colhead{} &\colhead{} & \colhead{} &\colhead{} 
&\colhead{} &\colhead{} &\colhead{Flux Density\tablenotemark{a}} &
\colhead{} &\colhead{} & \colhead{}
\\
\colhead{Object}   & \colhead{z$_{\rm em}$} 
&\colhead{$3.6\mu$m\tablenotemark{b}} & 
\colhead{$\sigma$(3.6$\mu$m)\tablenotemark{c}} & 
\colhead{$4.5\mu$m\tablenotemark{b}} & 
\colhead{$\sigma$(4.5$\mu$m)\tablenotemark{c}} & 
\colhead{$5.6\mu$m\tablenotemark{b}} & 
\colhead{$\sigma$(5.8$\mu$m)\tablenotemark{c}} & 
\colhead{$8.0\mu$m\tablenotemark{b}} & 
\colhead{$\sigma$(8.0$\mu$m)\tablenotemark{c}} & 
\colhead{24$\mu$m\tablenotemark{b}}   & 
\colhead{$\sigma$(24$\mu$m)\tablenotemark{c}} \\
\colhead{} & \colhead{} & 
\colhead{mJy} & \colhead{mJy} & 
\colhead{mJy} & \colhead{mJy} &
\colhead{mJy}  & \colhead{mJy} &
\colhead{mJy} & \colhead{mJy} & 
\colhead{mJy} & \colhead{mJy} \\
}
\startdata

BR J0006-6208	&4.51\tablenotemark{c} &0.316&0.001&0.256&0.002&0.345&0.012&0.605&0.013&3.34&0.13 \\
BR J0324-2918 	&4.63 &0.375&0.001&0.276&0.002&0.296&0.009&0.301&0.011&1.47&0.14 \\
SDSS J0836+0054	&5.82 &\ldots&\ldots&\ldots&\ldots&\ldots&\ldots&\ldots&\ldots&1.01&0.10\\
SDSS J0940+5848	&4.665&0.232&0.001&0.164&0.001&0.152&0.006&0.240&0.006&1.49&0.06 \\
SDSS J0941+5947	&4.82 &0.128&0.001&0.100&0.001&0.0981&0.006&0.160&0.005&0.60&0.05 \\
SDSS J1048+4637	&6.23 &\ldots&\ldots&\ldots&\ldots&\ldots&\ldots&\ldots&\ldots&0.86&0.09 \\
PSS J1140+6205	&4.51 &0.224&0.001&0.151&0.002&0.202&0.11&0.249&0.009&1.73 & 0.15 \\
SDSS J1148+5251	&6.41 &\ldots&\ldots&\ldots&\ldots&\ldots&\ldots&\ldots&\ldots&1.52&0.12 \\
BR 1202-0725	&4.694&0.461&0.001&0.375&0.001&0.410&0.001&0.657&0.002&2.37& 0.24 \\
PC 1247+3406	&4.897&0.142&0.001&0.118&0.001&0.110&0.006&0.168&0.006&0.81&0.14 \\
PSS J1715+3809	&4.52 &0.222&0.001&0.175&0.002&0.230&0.010&0.373&0.010&2.06&0.07 \\
PSS J1723+2243	&4.52 &1.197&0.003&0.969&0.003&1.062&0.010&1.286&0.010&6.21&0.15 \\
BR 2237-0607	&4.558&\ldots&\ldots&\ldots&\ldots&\ldots&\ldots&\ldots&\ldots&3.49&0.09 \\
\enddata

\tablenotetext{a}{The filters are sufficiently narrow that
uncertainties due to assumed SEDs and the color corrections are $<
5\%$ (IRAC \& MIPS Data Handbooks).  The conversions to mJy given in
SOM 4.6 are assumed here.}

\tablenotetext{b}{Wavelengths are the band designations for
each instrument.  The wavelengths corresponding to the monochromatic
flux density are listed in the SOM 4.6.  Band-passes are 21, 23, 25
and 36$\%$ for IRAC bands 3.6, 4.5, 5.8, and 8.0$\mu$m, and 20, 27 and
22$\%$ for MIPS 24, 70 and 160$\mu$m.}

\tablenotetext{c}{Measurement uncertainties only.  Absolute
calibration uncertainties are 3-5\% for IRAC (Reach et al.  2005) and
10\% for MIPS 24$\mu$m (SOM 4.6).}

\tablenotetext{c}{For our object selection and analysis, we adopt the discovery 
redshift from Storrie-Lombardi et al. (2001).  A slightly lower 
redshift ($z = 4.445$) is listed in Peroux et al. (2001).}
\end{deluxetable}

\end{document}